\documentclass[conference]{IEEEtran}
\IEEEoverridecommandlockouts
\usepackage{cite}
\usepackage{tikz}
\usetikzlibrary{arrows.meta,chains, positioning,shapes.geometric}
\usepackage{amsmath,amssymb,amsfonts}
\usepackage{algorithmic}
\usepackage{graphicx}
\usepackage{subcaption}
\usepackage{algorithm,algorithmic}
\usepackage{textcomp}
\usepackage{cellspace} 

\usepackage{makecell}
\usepackage[inline]{enumitem}
\usepackage{float}
\usepackage{soul}
\usepackage[normalem]{ulem}
\usepackage{xcolor}
\usepackage[font=footnotesize]{caption}
\usepackage{titlesec}
\usepackage{tikz}
\usetikzlibrary{shapes.geometric, arrows}
\usepackage{color,soul}
\usepackage{xcolor}
\usepackage{booktabs}
\usepackage{tabularx}
\usepackage{romannum}
\usepackage{balance}
\usepackage{mathtools}
\usepackage{soul}
\usepackage{scalerel}
\usepackage{diagbox}
\usepackage{multirow}
\usepackage{array}
\usepackage{threeparttable}
\usepackage{stackengine}
 \usepackage{caption}
\newcolumntype{P}[1]{>{\centering\arraybackslash}p{#1}}
\def\BibTeX{{\rm B\kern-.05em{\sc i\kern-.025em b}\kern-.08em
    T\kern-.1667em\lower.7ex\hbox{E}\kern-.125emX}}

\begin{document}
\setlength{\textfloatsep}{10pt}\setlength{\abovedisplayskip}{10pt}
\setlength{\belowdisplayskip}{10pt}
\title{Time of Arrival Error Estimation for Positioning Using Convolutional Neural Networks}

\author{\IEEEauthorblockN{Anil Kirmaz*$^\dagger$, Taylan \c{S}ahin*, Diomidis S. Michalopoulos*, Muhammad Ikram Ashraf*,\\ and Wolfgang Gerstacker$^\dagger$}
\IEEEauthorblockA{*Nokia Strategy and Technology, Munich, Germany, \\
$^\dagger$Institute for Digital Communications,
Friedrich-Alexander-Universität Erlangen-Nürnberg,
Erlangen, Germany}
%\\
%anil.kirmaz@fau.de}
}

\maketitle
\begin{abstract}

Wireless high-accuracy positioning has recently attracted growing research interest due to diversified nature of applications such as industrial asset tracking, autonomous driving, process automation, and many more. However, obtaining a highly accurate location information is hampered by challenges due to the radio environment. A major source of error for time-based positioning methods is inaccurate time-of-arrival (ToA) or range estimation. Existing machine learning-based solutions to mitigate such errors rely on propagation environment classification hindered by a low number of classes, employ a set of features representing channel measurements only to a limited extent, or account for only device-specific proprietary methods of ToA estimation. In this paper, we propose convolutional neural networks (CNNs) to estimate and mitigate the errors of a variety of ToA estimation methods utilizing channel impulse responses (CIRs). Based on real-world measurements from two independent campaigns, the proposed method yields significant improvements in ranging accuracy (up to 37\%) of the state-of-the-art ToA estimators, often eliminating the need of optimizing the underlying conventional methods.

\end{abstract}


%%%%%%%%%%%%%%%%%%%%%%%%%%%%%%%%%%%%%%%%%%%%%%%%%%%%%%%%%%%%%%%%%%%%%%%%%%%%%

\section{Introduction}
Location information is vital for many applications across various domains including industrial internet-of-things (IIoT), emergency services, transportation, and many more. Some of the applications, such as industrial asset tracking, autonomous driving and process automation, require highly accurate position estimation as emphasized in 3GPP [1]. Location information can be obtained by various approaches including time-based, angle-based and fingerprinting-based techniques using radio signals. One of the major approaches in widely utilized time-based positioning is to estimate the time-of-arrival (ToA) of the received positioning signals. Combined with time-of-transmission (ToT), i.e., the time when the radio signal is sent from the transmitter, ToA is used to calculate time-of-flight (ToF), i.e., the time it takes for the radio signal to travel from transmitter to receiver. Then, the range between transmitter and receiver can be estimated using ToF since the radio signals travel at a known speed, i.e., the speed of light, and can be utilized for positioning. \par

The accuracy of ToA estimation is limited by various factors such as challenging propagation conditions, synchronization errors, measurement inaccuracies and limitations in radio resources. Some of the factors, such as hardware properties and limited radio bandwidth, are determined strictly by the cost or regulation limitations and are more difficult to eliminate. However, some others that are related to the propagation environment may be detected and mitigated to some degree by convenient post-processing especially when a large radio bandwidth, e.g., that of an ultra-wideband (UWB) transmission, is available. Among propagation environment related factors, non-line-of-sight (NLOS) propagation is one of the primary error sources in time-based positioning methods since it de-correlates the time-of-flight (ToF) and the distance between transmitter and receiver. \par 

Various approaches have been proposed to improve the accuracy of the time-based positioning techniques through identifying or mitigating the effect of propagation conditions on positioning. Binary classification of the propagation environment has been studied commonly in the form of line-of-sight (LOS) versus non-line-of-sight (NLOS) classification using hypothesis testing based on probabilistic models [2], supervised machine learning (ML) [3], [4] and unsupervised ML [5], [6]. Furthermore, multi-class classification has been proposed by dividing NLOS propagation into two sub-classes depending on the partial or full blockage of the LOS path [7], [8], by adding a \textit{multipath} class to the binary classification problem [9], or by classifying the material of the LOS blocking objects [10]. \par

Even though the classification approach can improve the ranging or positioning accuracy through utilizing only the favorable, i.e., LOS, measurements [3], [4], [5], discarding NLOS measurements might lead to a poor positioning performance when the number of the available measurements is low. Moreover, such classification methods may not utilize the full information present in the measurements since the number of classes might be insufficient to describe the severity of the NLOS propagation in the measurements fully. \par

Ranging error mitigation by processing various features extracted from a received UWB waveform was studied by utilizing support vector machines and Gaussian process estimators [8], [11], or by fuzzy comprehensive evaluation along with propagation channel identification [12]. Although the methods were reported to yield an improvement in ranging, the predetermined features extracted from the received waveform might not represent all information in the received waveform with respect to the ranging error. Such information loss was overcome in [3], [13] where the ranging error was estimated directly from a given channel impulse response (CIR) by using artificial neural network (ANN) estimators. However, only a specific UWB measurement and ranging device (DWM1000 [14]) which utilizes a proprietary ranging algorithm was considered. Although a leading-edge detection method was mentioned to be used for ToA estimation in [14], details on the adopted detection algorithm were not provided. In [15], ToA estimation via convolutional neural networks (CNNs) was studied, and the corresponding performance was compared with that of some conventional, i.e., non-ML, ToA estimators. However, the CNNs were trained mainly with simulation data, and ToA \emph{error} estimation was not studied which can provide a measure of reliability of ToA estimation.   \par

In this paper, we investigate the problem of \emph{estimating the errors of various ToA estimators from a given CIR}. Then, the estimated errors can be mitigated to improve ranging accuracy and, thereby, performance of a positioning system. The main contributions of this paper are as follows:
\begin{itemize}
    \item We propose a novel CNN-based scheme to estimate and mitigate errors of various conventional ToA estimation algorithms with different computational complexity such as inflection point estimation (IFP) [16] and peak detection [17], and compare their performance to that of the leading-edge detection (LDE) [18] and the DWM1000 module [14] for a given CIR.
    \item We analyze the error mitigation performance of the proposed CNN estimator for the cases of optimized and suboptimal versions of the underlying ToA estimation algorithms.
    \item We evaluate the performance for two independent real-world datasets to ensure that the results are not specific or biased to a single measurement campaign. 
\end{itemize}
The analysis in this paper demonstrates that the proposed CNN-based error mitigation scheme improves the accuracy of the underlying conventional ToA estimators significantly even if they are improved with a basic error mitigation method. Furthermore, the proposed method is shown to provide a robust ranging performance in case the parameters of the underlying conventional ToA estimators are suboptimal. \par

\section{System Description}
The considered scheme is composed of a two-step process. In the first step, an initial ToA estimation is realized based on a given CIR by one of the conventional methods listed in Section~\Romannum{2}-B1. In the second step, the initial ToA estimate and the CIR are input to an ANN to estimate the error of the initial ToA estimation. Then, this information is utilized to mitigate the error of initial ToA estimation, according to
\begin{equation}
\widehat{\text{ToA}}' = \widehat{\text{ToA}}_{\text{conventional}} - \widehat{\epsilon}_{\text{ToA}},
\end{equation}
where $\widehat{\text{ToA}}_{\text{conventional}}$, $\widehat{\epsilon}_{\text{ToA}}$ and $\widehat{\text{ToA}}'$ represent the initial ToA estimated by a conventional method, the estimated error of the conventional ToA estimate and the mitigated ToA, respectively.

\subsection{CIR and ToA Estimation}
A CIR characterizes the communication channel and contains information on the travel time of radio signals from transmitter to receiver. Transmitted signals might arrive at the receiver from different paths, e.g., direct, reflected, or diffracted paths. ToA represents the arrival time of the first arriving signal at the receiver and can be determined from a given CIR. \par

\subsection{Baseline Methods}
\subsubsection{Conventional ToA Estimators}
In this work, we consider widely used conventional ToA estimators, namely Peak, IFP and LDE, as well as DWM:
\begin{itemize}
    \item \emph{Peak}: The delay time of the first peak of the CIR above a noise threshold is considered as ToA [17].
    \item \emph{IFP}: The delay time of the first point above a noise threshold where the CIR concavity changes [16] is estimated as ToA.
    \item \emph{LDE}: The CIR is filtered by a moving average window whose output is further passed through two different moving maximum window filters in parallel. The first delay time above a noise threshold where the output of the smaller maximum window filter exceeds the output of the larger maximum window filter by a factor, i.e., the leading-edge detection factor, is determined as ToA [18].
    \item \emph{DWM}: ToA is estimated by the DWM1000 device. The DWM estimates used in this paper are taken from the publicly available datasets [3], [19]. Although a leading-edge detection method was mentioned to be used for the  ToA estimation in the device's user manual [14], the details of the DWM1000’s internal estimation algorithm are not provided.
\end{itemize}

For Peak, IFP, and LDE, we define the noise threshold in terms of the relative path strength similar to [20], formulated as
\begin{equation}
\gamma_{th_i} = \alpha\: \text{max}\{\text{CIR}_i\} 
\end{equation}
with the noise threshold factor $\alpha$. LDE has three additional parameters, namely the leading-edge detection factor and the size of the small and large windows. The parameters of Peak, IFP and LDE are optimized by an exhaustive search to yield the lowest mean absolute ToA error. \par

\subsubsection{Benchmark ToA Error Mitigation Method}
In addition to the described conventional ToA estimators, we consider a benchmark scheme to estimate the \emph{error} of the ToA estimation conducted by these conventional methods. Denoted by \emph{CnstAvg}, this benchmark models the ToA error as constant and given by the mean of the error for each conventional ToA estimator. Following the estimation of ToA error, the error can be mitigated according to (1).

\subsection{Ranging Based on ToA Estimation}
The range, i.e., the distance between the tag and anchor, can be estimated by multiplying the mitigated ToA by the speed of the radio signals, i.e., speed of light, according to 
\begin{equation}
\widehat{R} = c(\widehat{\text{ToA}}' - \text{ToT}),
\end{equation}
where $c$ and $\widehat{R}$ represent the speed of light and the estimated range, respectively. ToT in (3) can be eliminated by using a two-way-ranging or a time-difference-of-arrival scheme. Subsequently, positioning of a target device can be performed by utilizing the range estimates with respect to multiple anchors with known locations. As a result, improving the accuracy of ToA estimates, i.e., through the error mitigation, yields an improved ranging, thereby, a more accurate positioning.

\tikzstyle{start} = [ellipse, minimum height=0.001cm, minimum width=0.001cm, text centered, draw=black, fill=yellow]
\tikzstyle{arrow} = [thick,->,>=stealth]
\tikzstyle{process} = [rectangle, minimum width=0.05cm, minimum height=0.03cm, text centered, draw=black, fill=green!]
\tikzstyle{process2} = [rectangle, minimum width=0.05cm, minimum height=0.03cm, text centered, draw=black, fill=cyan!]
\tikzstyle{output} = [rectangle, minimum width=0.05cm, minimum height=0.03cm, text centered, draw=white, fill=white!]
\tikzstyle{compute} = [circle, minimum size =0.001cm, text centered, draw=black, fill=white!]

\renewcommand{\thefootnote}{\alph{footnote}}
\renewcommand{\thefootnote}{\fnsymbol{footnote}}

\renewcommand{\thempfootnote}{\fnsymbol{mpfootnote}}

\vspace{0.3cm}

\begin{figure}
\begin{minipage}{\textwidth}
\begin{centering}
\begin{tikzpicture}[node distance=1.5cm]
\node (cir) [start] {\footnotesize CIR};
\node (bm) [process2, below of=cir, yshift=0.4cm] {\footnotesize \makecell{Conventional \\ ToA estimator}};
\node (toa_bm) [output, below of=bm, yshift=0.4cm] {\footnotesize $\widehat{\text{ToA}}$\textsubscript{conventional}\footnotemark};
\node (cnn) [process, left of=toa_bm, xshift=-1.4cm] {\footnotesize \makecell{CNN-based ToA error\\ mitigation scheme}};
\node (cnst) [process, right of=toa_bm, xshift=1.4cm] {\footnotesize \makecell{CnstAvg ToA error\\ mitigation scheme}};
\node (subtract_cnn) [compute, below of=cnn, yshift=+0.3cm] {-};
\node (subtract_cnst) [compute, below of=cnst, yshift=+0.3cm] {-};
\node (toa_cnn) [output, below of=subtract_cnn, yshift=0.65cm] {\footnotesize $\text{ToA}'$\textsubscript{conventional+CNN}\footnotemark};
\node (toa_cnst) [output, below of=subtract_cnst, yshift=0.65cm] {\footnotesize $\text{ToA}'$\textsubscript{conventional+CnstAvg}\footnotemark};

\draw [arrow] (cir) -- coordinate(bm-cir) (bm);
\draw [arrow] (bm) -- (toa_bm);
\draw[arrow] (cir)  -| (cnn);
\draw [arrow] (toa_bm) -- (cnn);
\draw [arrow] (cnn) -- node [text width=0.5cm,midway,left ] {\normalsize $\widehat{\epsilon}$\textsubscript{ToA}}  (subtract_cnn) ;
\draw [arrow] (cnst) -- node [text width=0.5cm,midway,right ] {\normalsize $\widehat{\epsilon}$\textsubscript{ToA}}  (subtract_cnst) ;
\draw[arrow] (toa_bm)  |- (subtract_cnn);
\draw[arrow] (toa_bm)  |- (subtract_cnst);
\draw [arrow] (subtract_cnn) -- (toa_cnn);
\draw [arrow] (subtract_cnst) -- (toa_cnst);
\end{tikzpicture}
\end{centering}
   \vspace{-0.3 cm}
\footnotetext[1]{\scriptsize Conventional ToA estimators: Peak, IFP, LDE, DWM.}
\footnotetext[2]{\scriptsize Proposed CNN-based error mitigation scheme: Conventional+CNN (e.g., Peak+CNN).}
\footnotetext[3]{\scriptsize Benchmark error mitigation scheme: Conventional+CnstAvg (e.g. LDE+CnstAvg).}
\end{minipage}
  \caption{Flow diagram and the naming of the considered ToA estimators.}
   \vspace{-0.1 cm}
\end{figure}

\section{Proposed Method}

\subsection{ToA Error Mitigation Using ANNs}
The complex nature of NLOS or multipath propagation poses a challenge to accurate modelling of ToA estimation error based on an input CIR. Therefore, an ANN seems a sensible choice to model the error of the ToA estimation. \par 

We employ a one-dimensional CNN similar to [3], [13] to estimate the error of the conventional ToA estimators based on the input CIR, since CNNs are shown to be useful in identifying spatial correlations among the input samples [21]. Besides the CIR, $\widehat{\text{ToA}}\textsubscript{\text{conventional}}$ is also input to the CNN. Then, the output of the CNN, $\widehat{\epsilon_{\text{ToA}}}$, is used to mitigate the error of the conventional ToA estimator according to (1). \par 

The utilized CNN comprises 3 convolutional layers followed by a fully connected layer. 16 output channels are used in each convolutional layer with a kernel size of 5 and a stride of 2 where no pooling layer is used in order to avoid a potential information loss. The rectified linear unit (ReLU) is used as the activation function in each neuron except for the output layer, and dropout regularization with a factor of 0.5 is utilized to prevent over-fitting. The CNNs are trained by using the Adam optimizer [22] with a learning rate of $10^{-3}$ and a batch size of 32 to minimize the mean-squared error (MSE) between the estimated and the real ToA error. \par 

The parameters of the CNN estimator are optimized using training and validation data. It was observed that increasing the number of hidden layers or number of output channels further does not result in a significant additional performance gain.

\subsection{Dataset Description and Pre-processing}
\subsubsection{Datasets}
We have used two publicly available datasets comprising real-world UWB measurements, which we refer to as \emph{Office} and \emph{Room}. Office dataset, given in [3], pertains to two different office environments, \emph{Office1} and \emph{Office2}. Room dataset, described in [13] and given in [19], comprises measurements taken in different sized office-like rooms with different dimensions. The measurements in both datasets are taken with 499.2 MHz of bandwidth at 3993.6 MHz of center frequency. \par

It is assumed that the propagation channel between transmitter and receiver is reciprocal, i.e., identical, for forward and backward transmit directions, and the channel coherence time is larger than the reply time of the applied two-way ranging system. Such assumptions are realistic and required since a single CIR is provided per each two-way ranging in the datasets. \par

 \begin{figure}[t]
\begin{center}
    \includegraphics[width=0.4\textwidth]{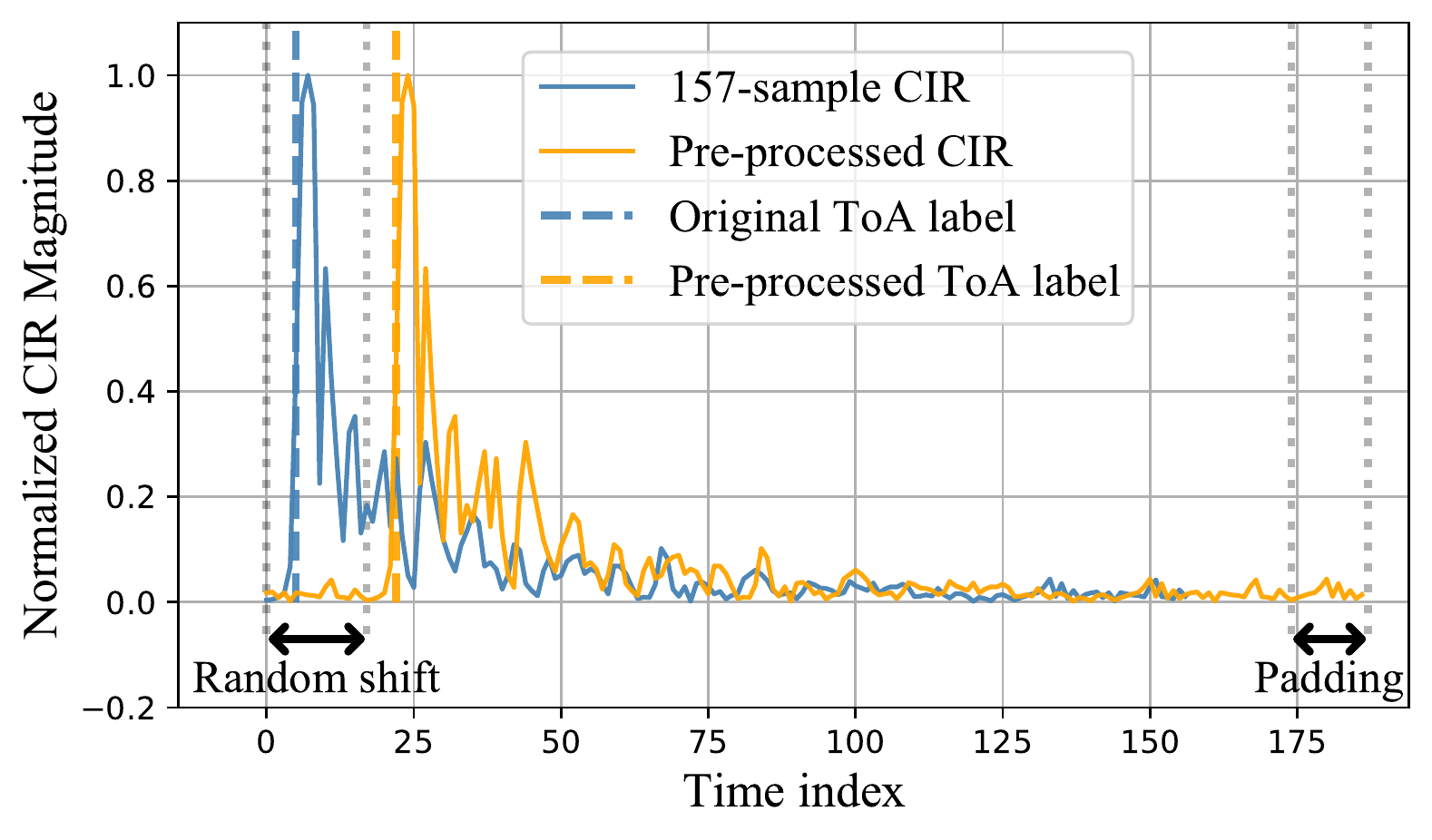}
\end{center}
    \vspace{-0.32 cm}
    \caption{Randomly shifted and padded CIRs using the described pre-processing.}
\label{fig:example}
\vspace{-0.25cm}
\end{figure}

\begin{figure*}
  %\centering
  \begin{subfigure}{.34\linewidth}
    \centering
    \includegraphics[width = .95\linewidth]{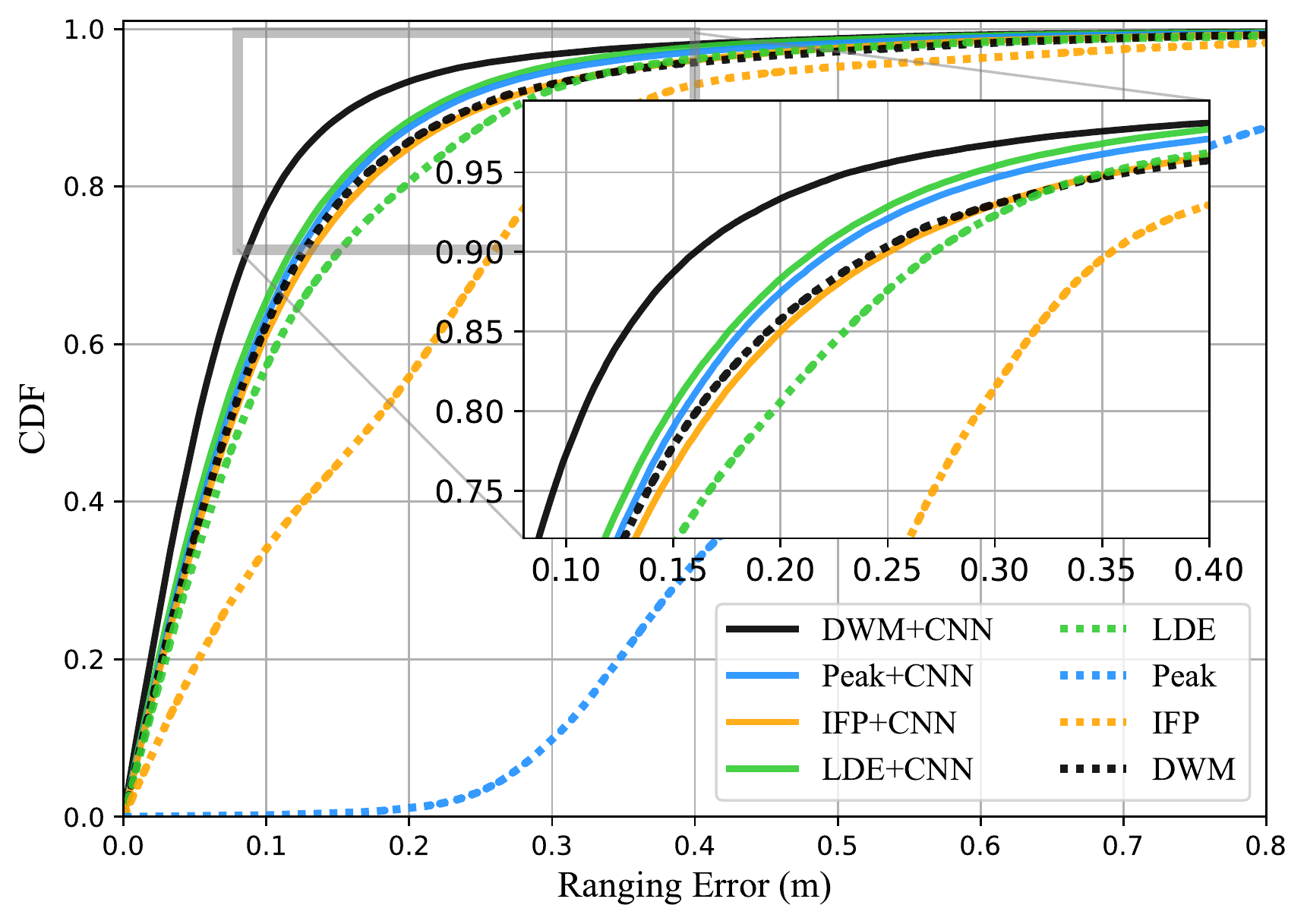}
         \vspace{-0.1 cm}
    \caption{}
  \end{subfigure}%
  %\hspace{2em}% Space between image A and B
  \begin{subfigure}{.34\linewidth}
   \centering
    \includegraphics[width = .95\linewidth]{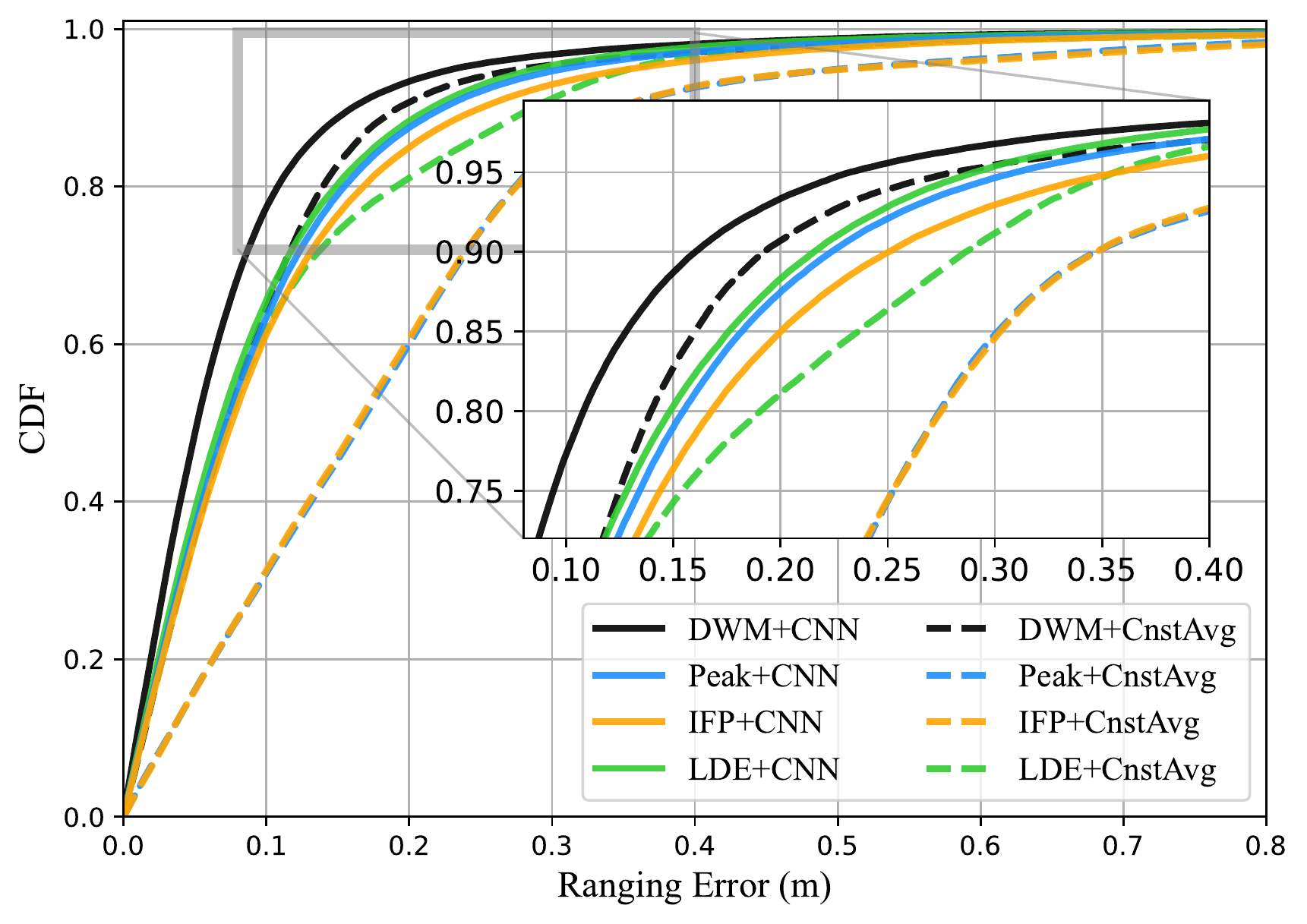}
               \vspace{-0.1 cm}
    \caption{}
  \end{subfigure}%
    \begin{subfigure}{.3\linewidth}
       \centering
    \includegraphics[width = 0.95\linewidth]{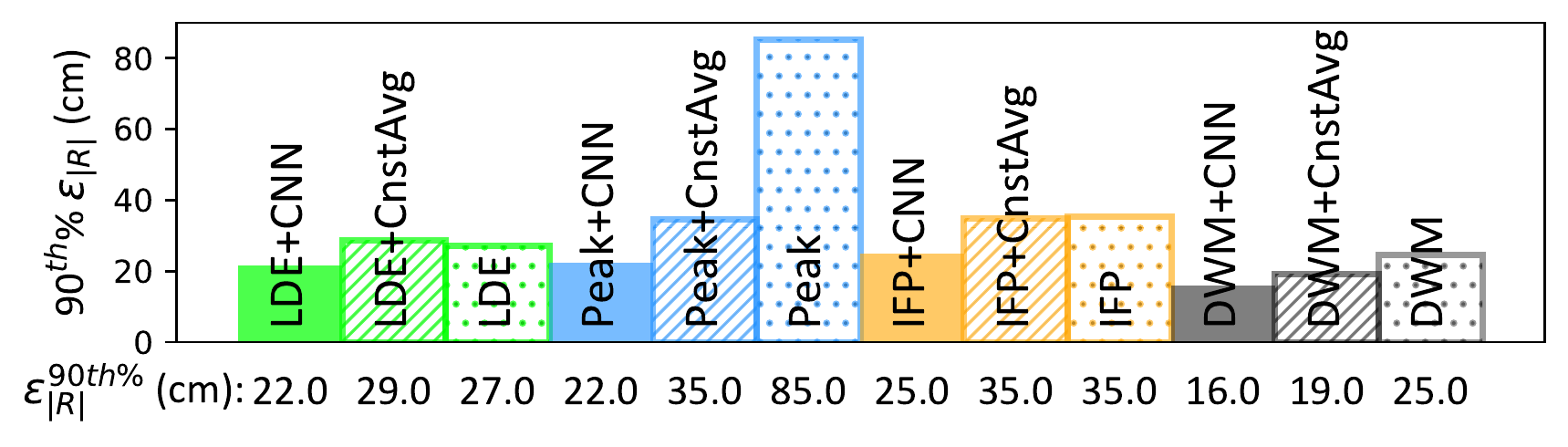}
             \vspace{-0.25 cm}
        \caption{}
                     \vspace{0.15 cm}

    \vfill
        \includegraphics[width = 0.95\linewidth]{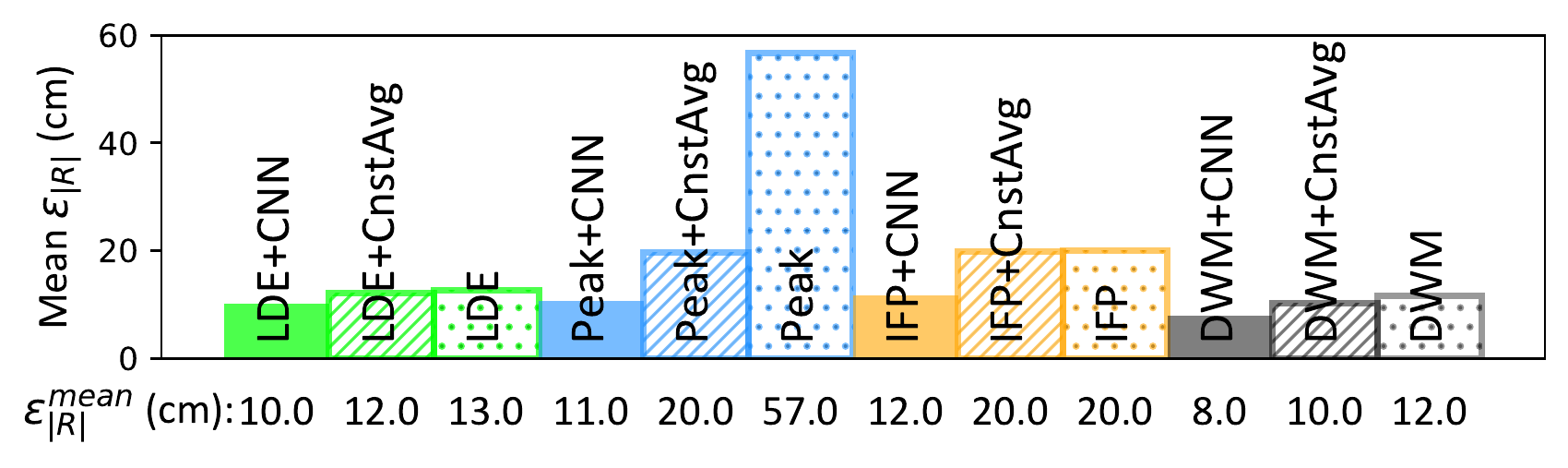}
         \vspace{-0.25 cm}
    \caption{}
      \end{subfigure}%
             \vspace{-0.1 cm}
%  \caption{The CDF of ranging error and gain at 80th percentile of the proposed CNN-based estimators in comparison to (a) conventional ToA estimators and  (b) benchmark CnstAvg estimators for Room dataset.}
    \caption{The CDF of ranging error of the proposed CNN-based estimator in comparison to (a) conventional ToA estimators and  (b) benchmark CnstAvg estimators, and comparison of  (c) 90th percentile and (d) mean absolute ranging error of the considered schemes} for Room dataset.
    
               \vspace{-0.55 cm}
\end{figure*}

\subsubsection{ToA Labeling}
The ToA delay time estimated by DWM, $\widehat{\text{ToA}}_{\text{DWM}}$, the corresponding ranging error, $\epsilon_R$, and time resolution of the CIR (i.e., the absolute time lapse between consecutive CIR indices), $\:\delta_t$, are given (or can be obtained) from the datasets [3], [19]. Utilizing this information, we determine the ground-truth ToA indices, i.e., ToA labels,  according to
\begin{equation}
\text{ToA}_{\text{true}} = \widehat{\text{ToA}}_{\text{DWM}} - \frac{\epsilon_R}{c\:\delta_t}. 
\end{equation}
As such, the ranging error is converted into a ToA error which is subtracted from the estimated ToA to determine the true ToA.

It should be noted that labeling real ToA in real-world CIR measurements is challenging and the introduced labeling may contain errors due to the clock drift, finite bandwidth and finite sampling rate.

\subsubsection{Data Pre-processing}
Only 152 (out of 1016) samples after the first detected path were considered for each CIR in [3], whereas additional 5 CIR samples prior to the detected first path were also considered in [13] yielding CIRs with 157 samples. We further add a random number of noise-like samples (maximum 30 samples) prior to each CIR shifting CIRs randomly with respect to the time axis to eliminate a potential bias, and apply padding to the end of CIRs accordingly, yielding CIRs with 187 samples as shown in Fig. 2. The ToA labels are shifted together, i.e., by the same amount, with the CIRs.

Each CIR is normalized by its maximum value before being input to the proposed CNN estimator to prevent a potential bias that might be caused by varying absolute amplitudes of the CIR samples. \par

The datasets are divided into training, validation and test data for the CNN. Further, to enable a fair comparison, the training and validation data are used together to optimize the parameters of the conventional ToA estimators and the benchmark error mitigation method. The test data is selected from measurements taken in another environment (i.e., another office or another sized-room) than the training and validation data to assess the generalizability of the results. This approach is in line with the recent 3GPP agreements on evaluating the generalization performance of ML models used for positioning [23]. Training and validation data comprise 70\% and 30\% of the measurements belonging to the same environment, respectively, resulting in approximately 5000 training samples in each scenario for the Office dataset. To make a fair comparison between the two datasets, we also use approximately 5000 training samples for each scenario in the Room dataset. It is noted that the Office dataset includes repeated measurements taken from each anchor-tag location pair, i.e., not all training samples is associated with a different anchor-tag location pair, unlike the Room dataset.

\section{Performance Evaluation}
In this section, we present performance results based on real-world measurements for the proposed (CNN) and the benchmark (CnstAvg) ToA error mitigation methods as well as conventional, i.e., unmitigated, ToA estimators (LDE, IFP, Peak, DWM). The naming of the estimators considered in this paper is shown in Fig. 1. We utilize the PyTorch framework to train the CNN. The results are generated based on 10 random selections of training and test measurement samples for each scenario to average out potential variations across data chunks.

\begin{figure*}
  %\centering
  \begin{subfigure}{.34\linewidth}
    \centering
    \includegraphics[width = .95\linewidth]{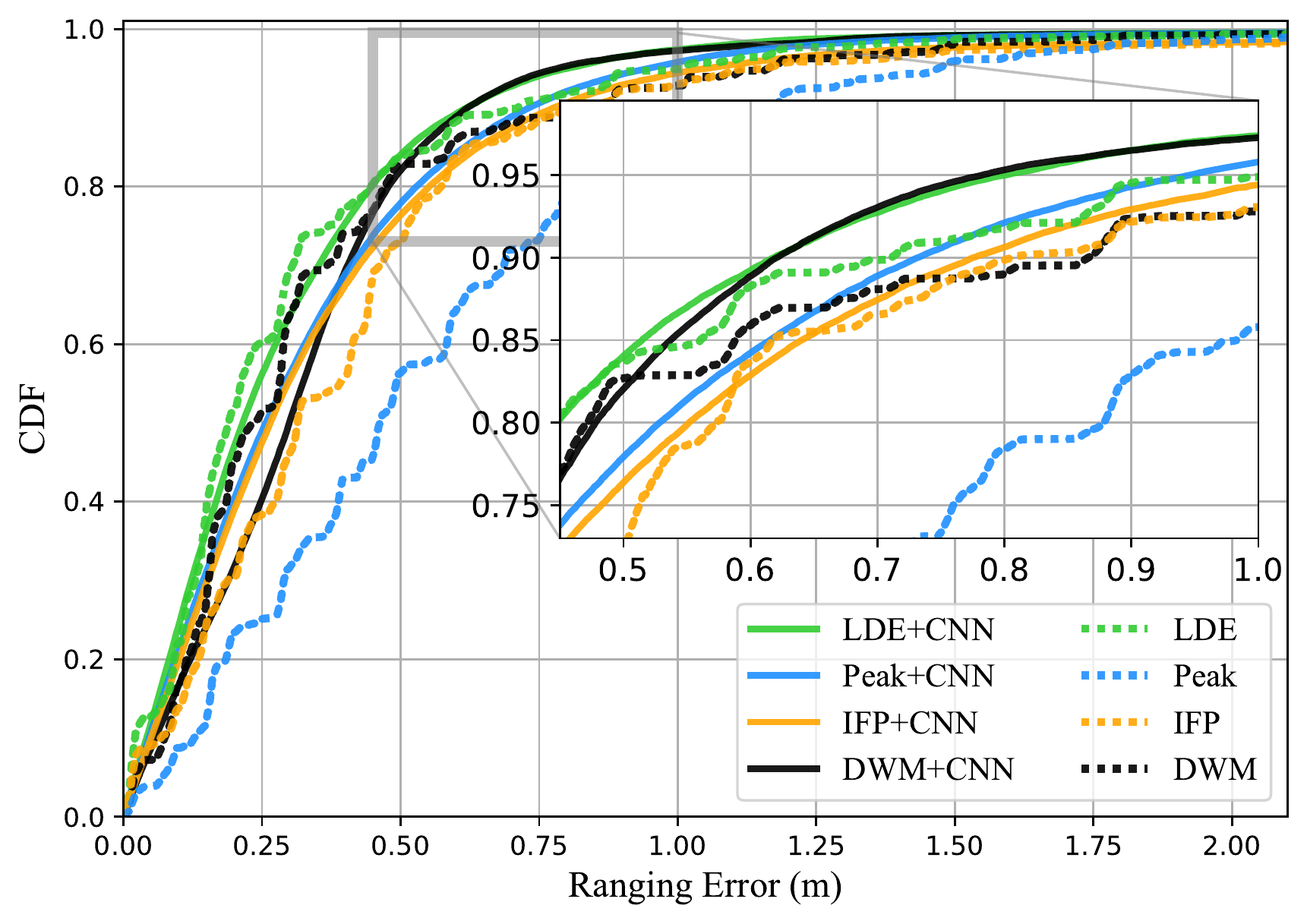}
         \vspace{-0.1 cm}
    \caption{}
  \end{subfigure}%
  %\hspace{2em}% Space between image A and B
  \begin{subfigure}{.34\linewidth}
   \centering
    \includegraphics[width = .95\linewidth]{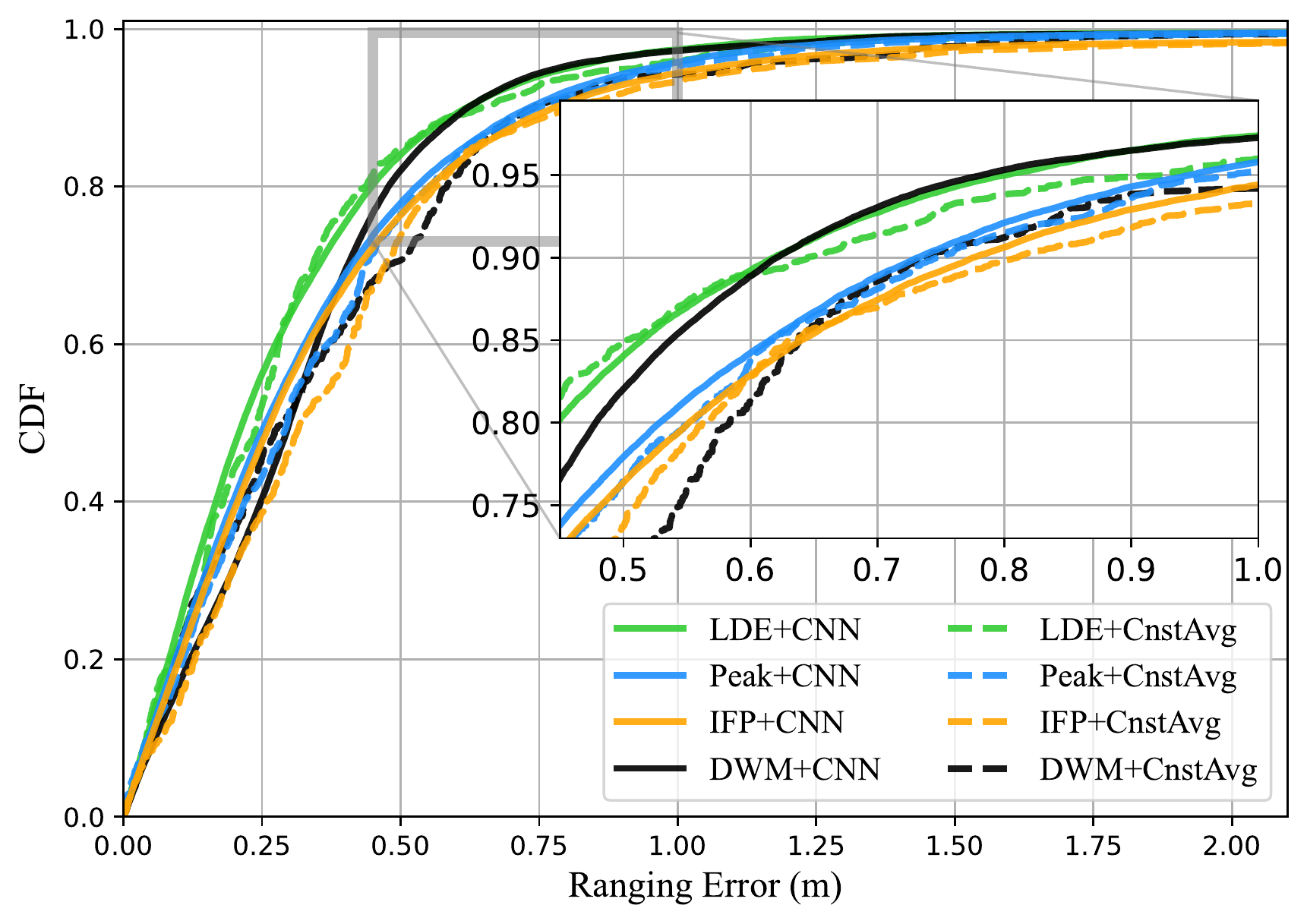}
               \vspace{-0.1 cm}
    \caption{}
  \end{subfigure}%
    \begin{subfigure}{.3\linewidth}
       \centering
    \includegraphics[width = 0.95\linewidth]{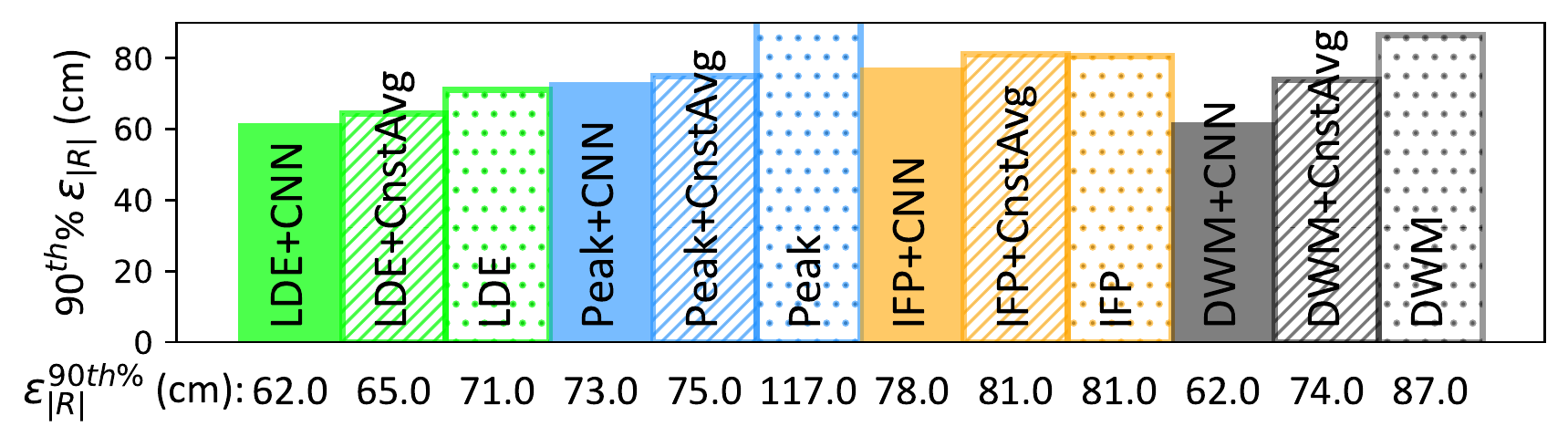}
             \vspace{-0.25 cm}
        \caption{}
                     \vspace{0.15 cm}

    \vfill
        \includegraphics[width = 0.95\linewidth]{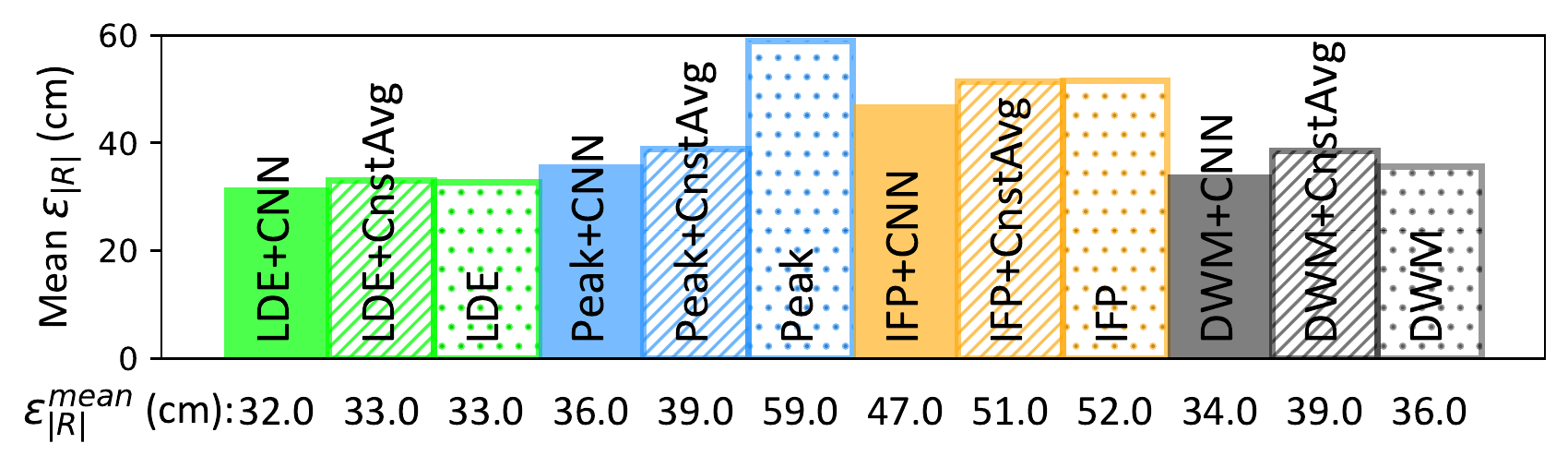}
         \vspace{-0.25 cm}
    \caption{}
      \end{subfigure}%
             \vspace{-0.1 cm}
    \caption{The CDF of ranging error of the proposed CNN-based estimator in comparison to (a) conventional ToA estimators and  (b) benchmark CnstAvg estimators, and (c) 90th percentile and (d) mean absolute ranging errors of the considered schemes} for Office dataset.
               \vspace{-0.55 cm}
\end{figure*}

\subsection{Ranging Accuracy Evaluation}

As evaluation metric, we consider the absolute ranging error, $\epsilon_{|R|}$, given by
\begin{equation}
\epsilon_{|R|} = |\Hat{R} - R_{\text{true}}| ,
\end{equation}
where $R_{\text{true}}$ denotes the real range obtained from the datasets [3], [19]. \par

We provide the CDF of the ranging error for different ToA estimation schemes in Figs. 3 and 4, for Office and Room datasets, respectively. It can be observed from Figs. 3-4 that the proposed CNN-based error mitigation scheme improves the accuracy of the conventional ToA estimators. The improvement in 90th percentile ranging error varies between 19-74\% and 4-38\% for Room and Office datasets, respectively, depending on the utilized conventional ToA estimator. The smaller improvement for the Office dataset can be explained by the fact that the Office dataset contains repeated measurements taken for the same anchor-tag location pairs, unlike the Room dataset. As a result, there is a lower number of measurements taken for \emph{unique} anchor-tag location pairs leading to an insufficient amount of unique data for the CNN to be trained. Furthermore, all methods perform worse in Office dataset than in Room dataset despite the same measurement and ranging module, DWM1000, used. This can be explained by the different propagation environments, i.e., the propagation environment for Office dataset might be more challenging, or a discrepancy in the calibration of the DWM1000 module, e.g., antenna delay calibration. \par

Comparing the two error mitigation methods, i.e., CNN and CnstAvg, the proposed CNN-based method further yields a considerably better performance than CnstAvg in most cases, and a similar performance in the worst case, depending on the underlying conventional ToA estimator. The gain of the CNN estimator over CnstAvg estimator lies between 16-37\% and 3-16\% in Room and Office datasets, respectively, in 90th percentile ranging accuracy. Our performance evaluation also enables a comparison of conventional ToA estimators from the literature. Figures 3a and 4a show that LDE outperforms IFP and Peak. Peak is observed to show the worst performance in both datasets possibly due to the susceptibility of the peak detection to multipath propagation [18], [24]. 

\subsection{Comparison with DWM}
 Figure 3a shows that DWM outperforms LDE slightly whereas LDE has a marginally better performance than DWM according to Fig. 4a. The similar performance of DWM and LDE can be explained by the fact that a leading-edge detection method was utilized by the DWM1000 device. Another observation is that CnstAvg degrades the performance of DWM, i.e., CnstAvg+DWM performs worse than DWM, in mean absolute ranging error for Office dataset. This can be explained by the fact that the average ToA estimation error of DWM is substantially different for Office1 and Office2, i.e., for training and test data. \par

Accuracy performance comparison of DWM+CNN and LDE+CNN shows contradicting results, similar to the comparison between DWM and LDE. LDE+CNN outperforms DWM+CNN for the Office dataset while DWM+CNN has the superior performance for the Room dataset. The underlying reason might be a discrepancy in the calibration of the DWM1000 device in the two measurement campaigns. The details of the DWM1000’s internal estimation algorithm were not provided neither in the device's user manual [14] nor in the descriptions of the measurements campaigns [3], [13]. Therefore, it is difficult to draw further conclusions regarding the performance of DWM-related estimators.

\subsection{Effect of Utilizing Sub-optimal Conventional Methods}
Various approaches can be used to optimize the parameters of the conventional methods. For instance, as an alternative to selecting the noise threshold in terms of the relative path strength [20], it can also be determined in terms of the thermal noise [11]. Additionally, the number and density of the candidate values of an exhaustive or grid search might yield different optimized parameters. As a result, the parameters of the utilized conventional ToA estimators can be sup-obtimal. \par 

In Table~\Romannum{1}, we provide the results related to the impact of optimizing the conventional ToA estimators. Such impact could not be evaluated for DWM since it is based on a proprietary detection algorithm. It can be observed from Table~\Romannum{1} that the performance of the conventional ToA estimators heavily depends on the parameter optimization for the measurements in both datasets. The proposed CNN estimator provides a robust ranging estimation, in case the utilized conventional ToA estimators are not optimized carefully. Specifically, using the proposed CNN estimator, the loss in ranging performance due to suboptimal parameters of the conventional ToA estimators is at most 8~cm at 90th percentile for both datasets, compared to 21~cm of CnstAvg and 37~cm of the conventional ToA estimators.

\subsection{Complexity Analysis}
Finding peaks of the input CIR dominates the computational complexity of Peak requiring $O(N)$ operations, where $N$ denotes the length of the CIRs. The complexity of  IFP is mainly determined by the calculation of the gradient where a subtraction and a division is performed for each element yielding a complexity of $O(N)$. LDE is composed of a moving average filter followed by two moving maximum filters where the outputs of the two moving maximum windows are compared element-wise. The window size is constant in all three filters, and the window is shifted through the CIR yielding an overall complexity of $O(N)$. \par

 \begin{table}[t]
\centering
\footnotesize
  \begin{threeparttable}
  \setlength\extrarowheight{0.4pt}
\caption{90th percentile absolute ranging errors of the considered ToA estimators and the increase in ranging error due to suboptimal (underlying) conventional ToA estimators.}
\centering
\begin{tabular}{ |c|c|c|c|c|}
%\begin{tabular}{ |Sc|Sc|Sc|Sc|Sc|Sc|Sc|}
 \hline
  \multicolumn{3}{|c|}{\multirow{2}{*}{ \makecell{ToA (error) \\ estimation method}}} & \multicolumn{2}{|Sc|}{90th\%$(\epsilon_{|R|})$ (cm)} \\
    \cline{4-5}
  \multicolumn{3}{|c|}{} & Office dataset & Room dataset\\
  \hline
    \multirow{9}{*}{\rotatebox[origin=c]{90}{\makecell{Ranging error when \\conventional estimators \\ are optimized}}} & \multirow{3}{*}{\rotatebox[origin=c]{90}{LDE}} & LDE & 
                                    71 & 27   \\  \cline{3-5}
    & & LDE+CnstAvg &                65 & 29 \\   \cline{3-5}
    & & LDE+CNN &                   62 & 22   \\   \cline{2-5}

    & \multirow{3}{*}{\rotatebox[origin=c]{90}{Peak}} & Peak 
                                    & 117 & 85   \\  \cline{3-5}
    &  & Peak+CnstAvg                   & 75 & 35   \\   \cline{3-5}
    & & Peak+CNN                     & 73 & 22    \\   \cline{2-5}
    
        & \multirow{3}{*}{\rotatebox[origin=c]{90}{IFP}} & IFP 
                                        & 81 & 35 \\  \cline{3-5}
    &  & IFP+CnstAvg                       & 81 & 35  \\   \cline{3-5}
    & & IFP+CNN                          & 78 & 25   \\     \cline{2-5}

   \hline

    \multirow{9}{*}{\rotatebox[origin=c]{90}{\makecell{Increase in ranging \\ error due to suboptimal \\ conventional estimator}}} & 
    
    \multirow{3}{*}{\rotatebox[origin=c]{90}{\makecell{LDE}}} & LDE 
                                    & +28 & +12  \\  \cline{3-5}
    & & LDE+CnstAvg                    &  +12 & +3  \\   \cline{3-5}
     & & LDE+CNN     & +8 & +0   \\  \cline{2-5}

         & \multirow{3}{*}{\rotatebox[origin=c]{90}{\makecell{Peak}}} & Peak 
                                    & +31 & +22  \\   \cline{3-5}
    &  & Peak+CnstAvg                   & +18 & +13  \\   \cline{3-5}
    & & Peak+CNN                     & +4 & +7  \\   \cline{2-5}

    & \multirow{3}{*}{\rotatebox[origin=c]{90}{\makecell{IFP}}} & IFP 
                                        & +37 & +34  \\   \cline{3-5}
    &  & IFP+CnstAvg                       & +11 & +21   \\   \cline{3-5}
    & & IFP+CNN                         & +6  & +6    \\   \cline{2-5}

   \hline
 \end{tabular}
 \vspace{-0.2 cm}
  \end{threeparttable}
 \end{table}

Each one-dimensional convolutional layer of the proposed CNN is associated to a constant filter size, and a constant number of filters is shifted along the input CIR. The subsequent single fully connected layer maps the output of the last convolutional layer to a scalar resulting in an overall complexity of $O(N)$.
Although the dependence of the complexity on the input CIR size is linear for the considered estimators, the complexities of the estimators are different. Table~\Romannum{2} shows the time complexity of inference of the estimators that are implemented using Pytorch, numpy and scipy libraries of Python programming language running on a computer equipped with Intel(R) Xeon(R) CPU E5-2680 v4 @ 2.40GHz and 24 GB of RAM. The additional latency caused by the proposed CNN-based error mitigation scheme is comparable to the latency of the widely used LDE estimator.

\begin{table}[t]
\centering
\caption{Computation time of the conventional estimators and the additional latency caused by the CNN mitigation scheme for one sample.}
\begin{tabular}{ |l|l|l|l|l|}
 \hline
    Estimator & Peak & IFP & LDE & \textcolor{blue}{+}CNN   \\
   \hline
    Time (ms) & 0.07 & 0.12 & 0.39 & \textcolor{blue}{+}0.35 \\
   \hline
 \end{tabular}
  \vspace{-0.2 cm}
 \end{table}

\section{Conclusions}
In this paper, we have proposed a supervised ML approach based on CNNs for estimation of the error of conventional ToA estimators. These estimates are in turn used for mitigating such errors to improve the ranging accuracy. We have evaluated the performance of the proposed methods using real-world measurements collected from various environments. We first observed that the performance of the conventional ToA estimators differ significantly from each other, and further require optimization of their parameters for an improved performance. While the errors of the conventional ToA estimators could be mitigated partly by a simple benchmark mitigation scheme, such approach might even result in a worse performance in some cases.\par

As an alternative, the proposed CNN-based error mitigation method can improve the ranging accuracy of the conventional ToA estimators with an acceptable amount of added latency. The proposed estimator was shown to outperform the benchmark error mitigation scheme by up to 16-37\% in 90th percentile ranging accuracy depending on the environment. In addition, it was shown that the proposed CNN estimator provides a robust ranging performance, with only less than 8 cm of additional ranging error in 90th percentile, in case the parameters of the underlying ToA estimators are suboptimal. Thus, the CNN estimator can eliminate the necessity of carefully optimizing the underlying conventional ToA estimators, depending on the accuracy requirements. In this way, the proposed method offers an attractive solution for improving the ranging accuracy, providing a robust performance under different conventional ToA estimation algorithms and across various propagation environments. \par 

In addition to the proposed use of ML to mitigate the error of conventional ToA estimators, ML methods can be also applied to estimate the ToA \emph{directly}, i.e., without requiring a conventional ToA estimator. Further research is needed to compare the performance of these two approaches.
\balance

%%%%%%%%%%%%%%%%%%%%%%%%%%%%%%%%%%%%%%%%%%%%%%%%%%%%%%%%%%%%%%%%%%%%%%%%%%%%%

%%%%%%%%%%%%%%%%%%%%%%%%%%%%%%%%%%%%%%%%%%%%%%%%%%%%%%%%%%%%%%%%%%%%%%%%%%%%%%%
\end{document}